\documentclass[12pt,reqno]{amsart}

\usepackage[T1]{fontenc}
\usepackage{lmodern}
\usepackage{amsmath,amssymb,amsthm,mathtools}
\usepackage{microtype}
\usepackage{placeins}
\usepackage{xcolor}
\usepackage[margin=1in]{geometry}
\usepackage{tikz}
\usetikzlibrary{arrows.meta,calc,decorations.markings,positioning}
\tikzset{
  zero set/.style={very thick,blue!65!black},
  guide/.style={thin,black!45,dashed},
  vector/.style={-{Stealth[length=2.4mm]},thick},
  phase arrows/.style={
    postaction={decorate},
    decoration={markings,
      mark=at position .15 with {\arrow{Stealth}},
      mark=at position .48 with {\arrow{Stealth}},
      mark=at position .81 with {\arrow{Stealth}}}}
}
\usepackage[colorlinks=true,linkcolor=blue!55!black,citecolor=green!38!black,urlcolor=blue!55!black]{hyperref}
\newcommand{\arXiv}[1]{\href{http://arxiv.org/abs/#1}{arXiv:#1}}

\allowdisplaybreaks
\numberwithin{equation}{section}

\newtheorem{theorem}{Theorem}[section]
\newtheorem{proposition}[theorem]{Proposition}
\newtheorem{lemma}[theorem]{Lemma}
\newtheorem{assumption}[theorem]{Assumption}
\theoremstyle{remark}

\newcommand{\R}{\mathbb R}

\newcommand{\Z}{\mathbb Z}
\newcommand{\T}{\mathbb T}
\newcommand{\cH}{\mathcal H}
\newcommand{\cL}{\mathcal L}
\newcommand{\cO}{\mathcal O}
\newcommand{\dd}{\,\mathrm d}
\newcommand{\ii}{\mathrm i}
\newcommand{\abs}[1]{\lvert #1\rvert}

\newcommand{\ang}[1]{\langle #1\rangle}
\DeclareMathOperator{\Ran}{Ran}
\DeclareMathOperator{\sgn}{sgn}

\title[Sharp Wannier decay]{On optimal Wannier decay in dimension three}
\author{Simon Becker}
\address{Bocconi University, Via Sarfatti 25, 20136 Milan, Italy}
\email{simon.becker@unibocconi.it}
\author{Zhongkai Tao}
\address{Institut des Hautes \'Etudes Scientifiques, 35 route de Chartres, 91440 Bures-sur-Yvette, France}
\email{ztao@ihes.fr}
\author{Mengxuan Yang}
\address{Texas A\&M University, College Station, TX 77843, USA}
\email{yangmx@tamu.edu}

\begin{document}

\begin{abstract}
For a Chern nontrivial Bloch bundle over the three torus, we construct an orthonormal Wannier basis with the optimal pointwise decay $\cO(\abs{x}^{-5/2})$. All but one of the generators decay exponentially. This resolves the question left open in \cite{BTY}.
\end{abstract}

\maketitle

\section{Introduction}

The topology of the Bloch bundle constrains the spatial localization of Wannier functions. In dimensions two and three, vanishing of the first Chern class is equivalent to the existence of an exponentially localized orthonormal Wannier basis \cite{Nenciu83,Panati,Brouder07,MPPT}. When the first Chern class is nonzero, the natural question is the optimal polynomial decay rate. Thouless \cite{Thouless84} conjectured the rate $\cO(\abs{x}^{-2})$ in dimension two. In \cite{BTY}, we proved this conjecture, including a full asymptotic expansion, and obtained the uniform rate $\cO(\abs{x}^{-7/3})$ in dimension three.

In this work, we establish the sharp pointwise decay rate $\cO(\abs{x}^{-5/2})$ in dimension three, improving the previously established $\cO(\abs{x}^{-7/3})$ decay rate and answering \cite[Question~1]{BTY}. For a rank $r$ Bloch bundle, the construction gives $r-1$ exponentially localized generators and one generator with the critical polynomial decay.

Let $\Gamma\subset\R^3$ be a lattice, with dual lattice $\Gamma^*$, and set $\cH=L^2(\R^3/\Gamma)$. Let $P=(P(k))_{k\in\R^3}$ be a family of orthogonal projections on $\cH$. Following \cite{BTY,Panati}, we make the following assumption.

\begin{assumption}\label{ass:A}
The rank of $P(k)$ is a fixed positive integer $r$. The map $k\mapsto P(k)$ is real analytic from $\R^3$ to $\cL(\cH , C^\infty(\R^3/\Gamma) )$ for $k\in\R^3$, and
\[
 P(k+p)=\tau(p)P(k)\tau(p)^{-1},\qquad
 \tau(p)f(u)=e^{-\ii p\cdot u}f(u),\qquad p\in\Gamma^*.
\]
\end{assumption}

The principal examples are spectral projections onto isolated families of Bloch bands for periodic Hamiltonians with smooth coefficients. The ranges $\Ran P(k)$ define a rank $r$ complex vector bundle $E\to\R^3/\Gamma^*$ through the identifications
\[
 (k,v)\sim(k+p,\tau(p)v),\qquad v\in\Ran P(k),\quad p\in\Gamma^*.
\]
This is the \emph{Bloch bundle} associated with $P$. A \emph{global (equivariant) Bloch frame} for $E$ is a family of measurable functions $\phi_a:\R^3\to\cH$, $a=1,\ldots,r$, satisfying
\[
 \phi_a(k+p)=\tau(p)\phi_a(k),\qquad p\in\Gamma^*,
\]
such that $\{\phi_1(k),\ldots,\phi_r(k)\}$ is an orthonormal basis of $\Ran P(k)$ for almost every $k$. The associated \emph{Wannier functions} are defined by the inverse Bloch transform:
\begin{equation}\label{eq:wannier}
 w_a(x)=\frac1{\abs{\R^3/\Gamma^*}}
 \int_{\R^3/\Gamma^*}e^{\ii k\cdot x}\phi_a(k,x)\dd k,
 \qquad a=1,\ldots,r.
\end{equation}
Here $\phi_a(k,x)$ is extended $\Gamma$-periodically in the spatial variable. The lattice translates
\[
 \mathcal W=\{w_a(\,\cdot-\gamma):\gamma\in\Gamma,\ a=1,\ldots,r\}
\]
form an orthonormal basis of the subspace associated with $P$ under the Bloch--Floquet transform; this is a \emph{Wannier basis} \cite{BTY,Kuchment16} with generators given by \eqref{eq:wannier}.

\begin{theorem}\label{thm:main}
Suppose Assumption~\ref{ass:A} holds and $c_1(E)\ne0$. Then there is an equivariant orthonormal Bloch frame $\phi_1,\ldots,\phi_r$ such that the corresponding Wannier functions $w_1,\ldots,w_{r-1}$ decay exponentially and
\[
 \abs{w_r(x)}\le C\ang{x}^{-5/2},
 \qquad \ang{x}:=(1+\abs{x}^2)^{1/2}.
\]
Moreover, no Wannier basis of a Chern nontrivial bundle can have every generator bounded by $C_a\ang{x}^{-\alpha}$ with $\alpha>5/2$.
\end{theorem}

The optimality assertion follows directly from the localization dichotomy of Monaco--Panati--Pisante--Teufel \cite{MPPT}. Indeed, if every generator satisfies $\abs{w_a(x)}\le C_a\ang{x}^{-\alpha}$ with $\alpha>5/2$, then
\[
 \int_{\R^3}\abs{x}^2\abs{w_a(x)}^2\dd x
 \le C+C\int_1^\infty t^{4-2\alpha}\dd t<\infty,
 \qquad a=1,\ldots,r.
\]
The localization dichotomy then forces $c_1(E)=0$, a contradiction.

The same theory gives a Bloch frame in $H^s_{\mathrm{loc}}(\R^3;\cH)^r$ for every $s<1$ \cite[Theorem~2.4]{MPPT}. Under the inverse Bloch transform, this corresponds to
\begin{equation}
    \label{eq:s-moment}
     \sum_{a=1}^r\int_{\R^3}\ang{x}^{2s}\abs{w_a(x)}^2\dd x<\infty,
 \qquad s<1.
\end{equation}
Thus the exponent $5/2$ is the pointwise decay rate at the weighted $L^2$ threshold.

\subsection*{Related works}
Wannier \cite{Wannier} first introduced an orthonormal real space representation of Bloch bands by Fourier transforming the Bloch states, which was later commonly referred to as Wannier functions. Kohn \cite{Kohn59} and Nenciu \cite{Nenciu83} proved exponential Wannier localization for isolated nondegenerate bands in one-dimensional and three-dimensional time-reversal periodic systems, respectively. Marzari--Vanderbilt \cite{MarzariVanderbilt,MarzariReview} introduced and studied composite Wannier functions for multiple bands. Panati \cite{Panati} and Brouder--Panati--Calandra--Mourougane--Marzari\ \cite{Brouder07} related exponential localization of composite Wannier bases to vanishing Chern classes in dimensions two and three. Kuchment \cite[Theorem~5.4]{Kuchment09} proved that sufficiently rapid polynomial localization of an orthonormal Wannier basis forces triviality of the Bloch bundle. In dimensions two and three, Monaco--Panati--Pisante--Teufel \cite{MPPT} showed that a Wannier basis with finite second moments exists (cf.~equation~\eqref{eq:s-moment}) if and only if the first Chern class vanishes. Recently, Li--Dong--Ledwith--Khalaf \cite{LiEtAl24} obtained $\mathcal{O}(\abs{x}^{-2})$ pointwise Wannier decay in two-dimensional Chern bands. See the survey \cite{Kuchment16} for more details. 

Allowing redundancy permits exponential localization even for nontrivial bundles. Kuchment \cite[Theorem~5.7]{Kuchment09} constructed finitely many exponentially localized generators whose translates form a Parseval frame. In dimension three, Auckly--Kuchment \cite{AucklyKuchment} showed that one additional generator suffices; related exponentially localized Parseval frames in the magnetic setting were constructed by Cornean--Monaco--Moscolari \cite{CMM}. 

Beyond periodic systems, Lu--Stubbs--Watson \cite{LuStubbsWatson} studied the existence and computation of generalized Wannier functions. Marcelli--Moscolari--Panati \cite{MarcelliMoscolariPanati} and Lu--Stubbs \cite{LuStubbs} proved that localization implies triviality without periodicity, and Rossi--Panati \cite{RossiPanati} gave a coarse geometric formulation in terms of Roe triviality in all dimensions.

\subsection*{Acknowledgment}
The authors thank Maciej Zworski and Ruixiang Zhang for discussions on Wannier localization and Fourier decay. M.Y. acknowledges support from the NSF grant DMS-2554813. S.B. acknowledges support from SNF Grant PZ00P2 216019. Z.T.~was partially supported by a grant from the National Research Agency (ANR) under the France 2030 program, reference ANR-25-CFFS-0012.

\subsection*{AI usage statement}
The key construction of the phase in \eqref{eq:modifier} was generated in a discussion with ChatGPT 5.6. In addition, we used ChatGPT 6 for editorial assistance, making the figures, and checking the estimates. The authors take responsibility for the mathematical content.

\section{Proof of Theorem~\ref{thm:main}}

We first recall the exponential decay of Wannier functions, which follows from the same contour deformation argument as in~\cite[Proposition 5.7]{notes}.

\begin{lemma}\label{lem:spatial-regularity}
Under Assumption~\ref{ass:A}, an analytic equivariant section has a pointwise exponentially decaying Wannier function.
\end{lemma}

As in \cite[Proposition~2.2 and Remark~3]{BTY}, we split off an analytic orthonormal $(r-1)$-frame and treat its orthogonal line bundle $L$, for which $c_1(L)=c_1(E)\ne0$. Lemma~\ref{lem:spatial-regularity} gives pointwise exponential decay of the first $r-1$ Wannier generators. We identify isomorphic Hermitian line bundles by smooth unitary isomorphisms. A fixed invertible linear change of coordinates preserves the asserted decay rates. By \cite[Lemma~4.1]{BTY}, we may therefore take
\[
 \Gamma=\Z^3,\qquad \Gamma^*=2\pi\Z^3,\qquad
 c_1(L)=m\left[\frac{dk_1}{2\pi}\wedge\frac{dk_2}{2\pi}\right].
\]
Hereafter $\T^d=(\R/2\pi\Z)^d$ and $\sigma=\sgn(m)=\pm1$.

\begin{proposition}\label{prop:helix}
There is a smooth section $s_0$ of $L$ whose zero set is the disjoint union of $|m|$ helices
\begin{equation}\label{eq:helix}
 \gamma_j(s)=\bigl(p_j+a(\cos s,\sin s),s\bigr),\qquad s\in\R/2\pi\Z,
\end{equation}
for distinct $p_j\in\T^2$ and sufficiently small $a\in(0,1/2)$. In tube coordinates
\[
 k_j(s,y)=\bigl(p_j+a(\cos s,\sin s)+y,s\bigr),\qquad y=(y_1,y_2),
\]
there is a smooth unit frame $e_j$ of $L$ such that, near $y=0$,
\begin{equation}\label{eq:germ}
 s_0(k_j(s,y))=(y_1+\ii\sigma y_2)e_j(s,y).
\end{equation}
\end{proposition}

\begin{figure}[htbp]
\centering
\begin{tikzpicture}[font=\small,line cap=round,line join=round]
  \begin{scope}[x={(0.68cm,0cm)},y={(0.27cm,0.17cm)},z={(0cm,0.68cm)}]
    \draw[black!45] (0,0,0)--(4,0,0)--(4,4,0)--(0,4,0)--cycle;
    \draw[black!45,dashed] (0,0,4)--(4,0,4)--(4,4,4)--(0,4,4)--cycle;
    \foreach \X/\Y in {0/0,4/0,4/4,0/4}
      \draw[black!45] (\X,\Y,0)--(\X,\Y,4);
    \draw[guide] (2,2,0)--(2,2,4);
    \draw[zero set] (2,2,0)--(2,2,4);
    \fill[blue!65!black] (2,2,0) circle (1.25pt);
    \fill[blue!65!black] (2,2,4) circle (1.25pt);
    \node[fill=white,inner sep=1pt] at (2.55,2,2.1) {$\{p\}\times\T$};
    \node[below] at (2,0,0) {$k_1$};
    \node[left] at (0.5,3.5,0) {$k_2$};
    \node[left] at (0,0,2.2) {$s$};
    \node[align=center] at (2,-1,-0.70) {vertical zero circle};
  \end{scope}

  \draw[-{Stealth[length=3mm]},very thick]
    (4.55,1.55)--(6.05,1.55)
    node[midway,above,align=center] {$s$-dependent\\translation};

  \begin{scope}[xshift=6.55cm,x={(0.68cm,0cm)},y={(0.27cm,0.17cm)},z={(0cm,0.68cm)}]
    \draw[black!45] (0,0,0)--(4,0,0)--(4,4,0)--(0,4,0)--cycle;
    \draw[black!45,dashed] (0,0,4)--(4,0,4)--(4,4,4)--(0,4,4)--cycle;
    \foreach \X/\Y in {0/0,4/0,4/4,0/4}
      \draw[black!45] (\X,\Y,0)--(\X,\Y,4);
    \draw[guide] (2,2,0)--(2,2,4);
    \draw[zero set]
      plot[domain=0:360,samples=120,variable=\t]
      ({2+0.72*cos(\t)},{2+0.72*sin(\t)},{4*\t/360});
    \fill[blue!65!black] ({2+0.72},2,0) circle (1.25pt);
    \fill[blue!65!black] ({2+0.72},2,4) circle (1.25pt);
    \node[fill=white,inner sep=1pt] at (2.95,2.20,2.15) {$\gamma(s)$};
    \node[below] at (2,0,0) {$k_1$};
    \node[left] at (0.5,3.5,0) {$k_2$};
    \node[left] at (0,0,2.2) {$s$};
    \node[align=center] at (2,-1,-0.70) {helical zero circle};
  \end{scope}
\end{tikzpicture}
\caption{One component of the zero set in a fundamental domain of $\T^3$. Pullback from $\T^2$ gives a vertical circle; an $s$-dependent translation in the transverse variables bends it into $\gamma(s)=(p+a(\cos s,\sin s),s)$.  The endpoints at $s=0$ and $s=2\pi$ are identified.}
\label{fig:helix-construction}
\end{figure}
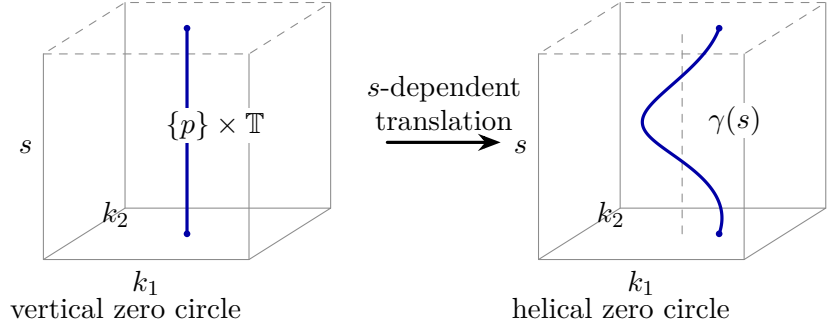

\begin{proof}
Let $\pi:\T^3\to\T^2$ be the projection onto the first two variables. The classification of complex line bundles by their first Chern class \cite[Chapter 17, Theorem 3.4]{Husemoller} gives $L\simeq\pi^*L_0$, where $L_0\to\T^2$ has Chern number $m$.

 Choose disjoint discs $D_j$ centered at $p_j$, $j=1,\ldots,|m|$. Choose a smooth section that vanishes at $p_j$ and has index $\sigma$ at each point $p_j$. Then we may glue the section $y_1+\ii\sigma y_2$ into the discs $D_j$ as in \cite[Proposition 4.2]{BTY}.

Finally, pull back by
\[
 F(k_1,k_2,s)=(k_1-a\cos s,k_2-a\sin s,s).
\]
Since $F$ is isotopic to the identity, $F^*L\simeq L$. After a smooth unitary identification, the pullback section has zero set \eqref{eq:helix} and local form \eqref{eq:germ}; see Figure~\ref{fig:helix-construction}.
\end{proof}

The normalized section $\phi^{(0)}=s_0/\|s_0\|$ is smooth away from the helices. On each punctured tube,
\begin{equation}\label{eq:vortex}
 \phi^{(0)}(k_j(s,r\cos\theta,r\sin\theta))
 =e^{\ii\sigma\theta}e_j(s,r\cos\theta,r\sin\theta).
\end{equation}

In order to get the optimal decay rate, we need to modify this section by a phase factor. Write
\[
 x=n+u,\qquad n\in\Z^3,\quad u\in\Omega:=[0,1)^3,\qquad
 x_\perp=(x_1,x_2)=\rho\omega_\beta,\quad \omega_\beta=(\cos\beta,\sin\beta).
\]
In a local lift of the bundle, $e_j(s,y)(u)$ denotes evaluation at the spatial variable $u$. Its derivatives in $s$ and $y$ are uniformly bounded in $u$.

\begin{lemma}\label{lem:transverse}
Let $H_s(\theta)$ be smooth in $(s,\theta)$ and satisfy $H_s(\theta+\pi)=-H_s(\theta)$. Let $\chi(|y|)$ be a smooth radial cutoff equal to one near zero and supported inside the tube. For $\rho\ge0$, let
\[
 T_{j,H}(s,\rho,\beta;u)=\int_{\R^2}e^{\ii\rho\omega_\beta\cdot y}
 H_s(\arg y)\chi(|y|)e_j(s,y)(u)\dd y.
\]
For $\rho\ge1$,
\begin{equation}
\label{eq:transverse}
 \begin{gathered}
 T_{j,H}(s,\rho,\beta;u)
 =\rho^{-2}C_H(s,\beta)e_j(s,0)(u)+R_{j,H}(s,\rho,\beta;u),\\
 \sup_{s,\beta,u}|\partial_s^N R_{j,H}(s,\rho,\beta;u)|
 \le C_N\rho^{-3},\qquad N\ge0,
 \end{gathered}
\end{equation}
where
\begin{equation}\label{eq:one-derivative}
 C_H(s,\beta)=-2\pi\ii\,\partial_\theta H_s(\beta+\pi/2).
\end{equation}
\end{lemma}

\begin{proof}
Replacing $e_j(s,y)(u)$ by $e_j(s,0)(u)$ introduces an amplitude vanishing to first order at $y=0$. Integration by parts on dyadic annuli $|y|\simeq 2^{-j}$ bounds its Fourier transform, also after any fixed number of $s$ derivatives, by
\[
 C_M\sum_{j}2^{-3j}(1+2^{-j}\rho )^{-M}
 \le C\rho^{-3},\qquad M>3.
\]
Removing $\chi$ from the homogeneous leading term contributes $\cO(\rho^{-\infty})$ as in \cite{BTY}. Scaling the radial variable therefore gives \eqref{eq:transverse}, with
\[
 C_H(s,\beta)=\lim_{\varepsilon\to 0^{+}}
 \int_0^{2\pi}\int_0^\infty
 e^{-\varepsilon r}e^{\ii r\cos(\theta-\beta)}H_s(\theta)r\dd r\dd\theta.
\]
Since
\[
 \int_0^\infty e^{-\varepsilon r}e^{\ii cr}r\dd r=-(c+\ii\varepsilon)^{-2},
 \qquad H_s(\theta+\pi)=-H_s(\theta),
\]
we have
\[
 C_H(s,\beta)=-\lim_{\varepsilon\downarrow0}\int_\beta^{\beta+\pi}
 H_s(\theta)\bigl[(\cos(\theta-\beta)+\ii\varepsilon)^{-2}
 -(\cos(\theta-\beta)-\ii\varepsilon)^{-2}\bigr]\dd\theta.
\]
Using
\[
 (t+\ii0)^{-2}-(t-\ii0)^{-2}=2\pi\ii\,\delta'(t),
\]
and the unique zero of $\cos(\theta-\beta)$ in $(\beta,\beta+\pi)$, we obtain
\[
 C_H(s,\beta)=-2\pi\ii\int_\beta^{\beta+\pi}
 H_s(\theta)\delta'(\cos(\theta-\beta))\dd\theta
 =-2\pi\ii\,\partial_\theta H_s(\beta+\pi/2).
\]
This proves \eqref{eq:one-derivative}.
\end{proof}

\begin{figure}[ht]
\centering
\begin{tikzpicture}[font=\small,line cap=round,line join=round]
  \begin{scope}[xshift=3.00cm]
    \draw[guide] (0,0) circle (1.35);
    \draw[vector] (0,0)--(2.25,0)
      node[below=4pt] {$\omega_\beta$};
    \node[below left=3pt] at (0,0) {$p$};
    \draw[thick] (-0.55,1.35)--(0.55,1.35);
    \draw[thick] (-0.55,-1.35)--(0.55,-1.35);
    \fill[blue!65!black] (0,1.35) circle (1.8pt);
    \fill[blue!65!black] (0,-1.35) circle (1.8pt);
    \node at (0,1.94)
      {$t=\pi/2:\ \psi''=0,\ \cos^2 t=0$};
    \node at (0,-1.94)
      {$t=-\pi/2:\ \psi''=0,\ \cos^2 t=0$};
    \node[align=center] at (0,-2.85)
      {$a(\cos s,\sin s)$\\(after rotating $\beta$ to $0$)};
  \end{scope}

  \begin{scope}[xshift=10.90cm]
    \draw[thin,black!55,-{Stealth[length=1.7mm]}]
      (-3.57,-1.63)--(3.65,-1.63);
    \draw[thin,black!55,-{Stealth[length=1.7mm]}]
      (-3.57,-1.63)--(-3.57,1.60);
    \draw[thin,black!30] (-3.57,0)--({180/55},0);
    \foreach \t in {-90,90}
      \draw[guide] ({\t/55},-1.63)--({\t/55},1.43);
    \draw[black,dashed,thick]
      plot[domain=-180:180,samples=181,variable=\t]
      ({\t/55},{-1.35*cos(\t)});
    \draw[blue!65!black,very thick]
      plot[domain=-180:180,samples=181,variable=\t]
      ({\t/55},{1.35*cos(\t)^2});
    \foreach \t/\lab in {-180/{-\pi},-90/{-\pi/2},0/{0},90/{\pi/2},180/{\pi}}
      \draw[thin,black!55] ({\t/55},-1.59)--({\t/55},-1.67)
        node[below=4pt,text=black] {$\mathstrut\lab$};
    \foreach \v in {-1,0,1}
      \draw[thin,black!55] (-3.61,{1.35*\v})--(-3.53,{1.35*\v})
        node[left=5pt,text=black] {$\v$};
    \draw[black,dashed,thick] (-3.48,1.94)--(-2.93,1.94);
    \node[anchor=west,inner xsep=0pt] at (-2.80,1.94)
      {$\psi''/(a\rho)=-\cos t$};
    \draw[blue!65!black,very thick] (1.32,1.94)--(1.87,1.94);
    \node[anchor=west,inner xsep=0pt] at (2.00,1.94) {$\cos^2 t$};
    \node at (0,-2.85) {$t=s-\beta$};
  \end{scope}
\end{tikzpicture}
\caption{The factor $\cos^2(s-\beta)$ vanishes quadratically at both zeros of $\psi''(s)$. At these points, $\psi'(s)=0$ precisely when $x_3=\pm a\rho$, giving a degenerate stationary point.}
\label{fig:fold-cancellation}
\end{figure}
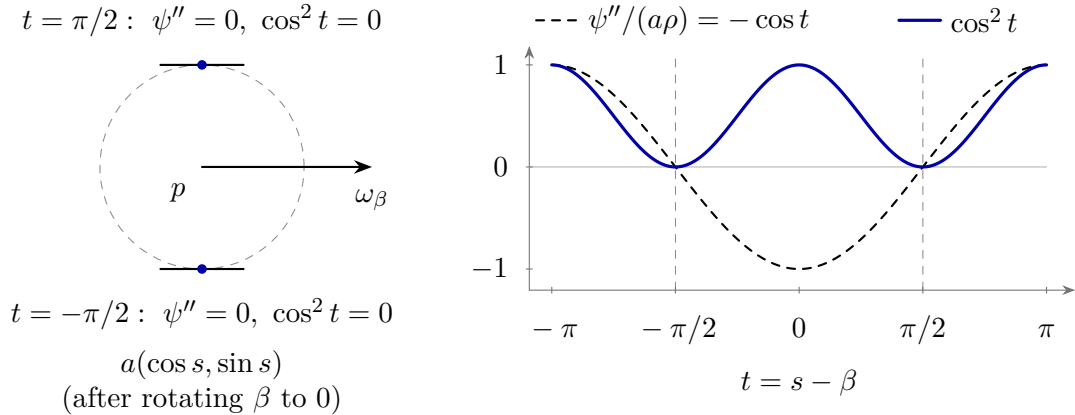

The longitudinal phase along a helix is
\begin{equation}\label{eq:longitudinal-phase}
 \psi(s)=a\rho\cos(s-\beta)+sx_3,\qquad
 \psi''(s)=-a\rho\cos(s-\beta).
\end{equation}
For $H_s(\theta)=e^{\ii\sigma\theta}$, the coefficient \eqref{eq:one-derivative} does not vanish where $\psi''=0$. A cubic stationary point then gives the factor $\rho^{-1/3}$ and the total decay rate $\rho^{-7/3}$ of~\cite{BTY}. We choose the following angular phase:
\begin{equation}\label{eq:H}
 \begin{aligned}
 f_s(\theta)=\theta-\tfrac12\sin2(\theta-s), \quad
 f_s'(\theta)=2\sin^2(\theta-s), \quad 
 H_{\sigma,s}(\theta)=e^{\ii\sigma f_s(\theta)}.
 \end{aligned}
\end{equation}
Note that $f_s(\theta+\pi)=f_s(\theta)+\pi$, and the winding number remains $\sigma$. Evaluating the derivative at $\theta=\beta+\pi/2$ yields
\begin{equation}\label{eq:C}
 C_{\sigma,s}(\beta)=4\pi\ii\cos^2(s-\beta)
 e^{\ii\sigma\beta+\frac{\ii\sigma}{2}\sin2(\beta-s)}.
\end{equation}
Thus $C_{\sigma,s}(\beta)=0$ at the zeros of $\psi''$; see Figure~\ref{fig:fold-cancellation}. The quadratic vanishing will give a uniform bound $\cO(\rho^{-1/2})$, hence the total decay rate $\rho^{-5/2}$.

\FloatBarrier

To realize this profile globally, choose $\eta\in C_c^\infty([0,\infty))$ equal to one near zero and supported inside the tubes. Multiply \eqref{eq:vortex} on each punctured tube by
\begin{equation}\label{eq:modifier}
 g(s,r,\theta)=\exp\!\left(-\frac{\ii\sigma}{2}\eta(r)\sin2(\theta-s)\right).
\end{equation}
Since $g=1$ near the tube boundary, this defines a unit section $\phi$ smooth away from the helices, with angular profile \eqref{eq:H} near each core. 

Choose $\chi$ as in Lemma~\ref{lem:transverse} supported in $\{\eta=1\}$, and let $\chi_j(k_j(s,y))=\chi(|y|)$. The inverse Bloch transform gives
\begin{equation}\label{eq:tube-integral}
 \begin{split}
 (2\pi)^3w_r(x)&=\sum_j I_j(x)+\cO(\ang{x}^{-\infty}),\\
 I_j(x)&=e^{\ii\rho\omega_\beta\cdot p_j}
 \int_0^{2\pi}e^{\ii\psi(s)}T_j(s,\rho,\beta;u)\dd s,
 \end{split}
\end{equation}
where $T_j=T_{j,H_{\sigma,\cdot}}$. We use the following form of the van der Corput lemma \cite[Chapter~VIII, \S1]{Stein}.

\begin{lemma}\label{lem:vdC2}
Let $I$ be a bounded interval, let $\varphi\in C^2(\overline I)$ satisfy $|\varphi''|\ge\Lambda>0$ and let $b\in C^1(\overline I)$. Then
\begin{equation}\label{eq:vdC2}
 \left|\int_Ie^{\ii\varphi(s)}b(s)\dd s\right|
 \le C\Lambda^{-1/2}\bigl(\|b\|_{L^\infty(I)}+\|b'\|_{L^1(I)}\bigr).
\end{equation}
\end{lemma}

\begin{proof}[Proof of the decay bound in Theorem~\ref{thm:main}]
We may assume $|x|\ge2$. For the phase in \eqref{eq:longitudinal-phase},
\[
 \psi'(s)=-a\rho\sin(s-\beta)+x_3.
\]
We split into two regimes.

\noindent\emph{Case 1.} Suppose $|x_3|>\rho$. We integrate by parts in $s$ in the full tube integral \eqref{eq:tube-integral}, before applying Lemma~\ref{lem:transverse}. Since $a<1/2$,
\[
 |\psi'(s)|\ge(1-a)|x_3|\gtrsim|x|,\qquad
 |\partial_s^\ell(1/\psi')|\le C_\ell|x|^{-1}.
\]
All $s$ derivatives of $H_{\sigma,s}(\arg y)\chi(|y|)e_j(s,y)(u)$ are bounded uniformly in $y$ and $u$. The integrand has matching endpoint values: equivariance gives
\[
 e_j(s+2\pi,y)(u)=e^{-2\pi\ii u_3}e_j(s,y)(u),\qquad
 e^{2\pi\ii x_3}e^{-2\pi\ii u_3}=e^{2\pi\ii n_3}=1.
\]
The same cancellation holds after differentiation. Each integration by parts therefore gives one power of $|x|^{-1}$, and
\begin{equation}\label{eq:case-1}
 |w_r(x)|\le C_N\ang{x}^{-N},\qquad N\ge0.
\end{equation}

\noindent\emph{Case 2.} Suppose $|x_3|\le\rho$, so $\rho\ge1$ and $\rho\asymp|x|$. Lemma~\ref{lem:transverse} and \eqref{eq:C} give
\begin{equation}\label{eq:ibp-1}
 I_j(x)=4\pi\ii\rho^{-2}e^{\ii\rho\omega_\beta\cdot p_j}
 \int_0^{2\pi}a_j(s,\beta;u)\cos^2(s-\beta)e^{\ii\psi(s)}\dd s
 +\cO(\rho^{-3}),
\end{equation}
where
\[
 a_j(s,\beta;u)=e^{\ii\sigma\beta+\frac{\ii\sigma}{2}\sin2(\beta-s)}e_j(s,0)(u).
\]
This amplitude is independent of $\rho$, and its $s$ derivatives are uniformly bounded in $\beta$ and $u$.

Away from the two zeros of $\cos(s-\beta)$, Lemma~\ref{lem:vdC2} gives $\cO(\rho^{-1/2})$. Near either zero $q=\beta+(k+1/2)\pi$, use the dyadic intervals
\[
 I_{\ell,q}^{\pm}=q\pm[2^{-\ell-1},2^{-\ell}],\qquad \ell\ge0,
\]
in a local lift of the $s$ circle. With $b(s)=a_j(s,\beta;u)\cos^2(s-\beta)$, we have
\[
 |\psi''|\gtrsim\rho\,2^{-\ell},\qquad
 \|b\|_{L^\infty(I_{\ell,q}^{\pm})}
 +\|b'\|_{L^1(I_{\ell,q}^{\pm})}\le C2^{-2\ell}.
\]
Thus
\[
 \left|\int_{I_{\ell,q}^{\pm}}b(s)e^{\ii\psi(s)}\dd s\right|
 \le C(\rho\,2^{-\ell})^{-1/2}2^{-2\ell}
 =C\rho^{-1/2}2^{-3\ell/2}.
\]
Summing over these intervals and the complementary intervals yields
\[
 \left|\int_0^{2\pi}a_j(s,\beta;u)\cos^2(s-\beta)e^{\ii\psi(s)}\dd s\right|
 \le C\rho^{-1/2}.
\]
Consequently, \eqref{eq:tube-integral} and \eqref{eq:ibp-1} imply
\[
 |w_r(x)|\le C\rho^{-2}\rho^{-1/2}+C\rho^{-3}
 \le C\ang{x}^{-5/2}.
\]
Together with Case 1 and the exponential decay of the other generators, this proves the theorem.
\end{proof}

\end{document}